\documentstyle[12pt,equations]{article}
\def\9{\phantom 0}      
\renewcommand\linebreak{\unskip\break} 

\def\lsim{\lower3pt\hbox{$\buildrel<\over\sim$}}

\def\to{\rightarrow}

\begin{document}
\input psfig.sty
\newlength{\captsize} \let\captsize=\small 
%
\rightline{OCIP/C-95-2}
\rightline{ UPR-648-T}
\rightline{hepph-9504216}
\rightline{March  1995}

\begin{center}
{\bf \large
DISCOVERY AND IDENTIFICATION OF \\[3pt]
EXTRA GAUGE BOSONS}
\footnote{Summary of the Working Subgroup on Extra Gauge Bosons of the
DPF long-range planning study to be published in
{\it   Electro-weak Symmetry Breaking and Beyond the Standard
Model}, eds T. Barklow, S. Dawson, H. Haber and
J. Seigrist (World Scientific 1995).}
\rm
\vskip1pc
Mirjam Cveti\v c \\
{\em Department of Physics, University of Pennsylvania \\
Philadelphia, PA~19104-6396}\\
\vspace{3pt}
and\\
\vspace*{3pt}
Stephen Godfrey \\
{\em Department of Physics, Carleton University \\
Ottawa, ON~K1S 5B6}
\end{center}
\vskip2pc
\begin{abstract}
The discovery potential and diagnostic abilities of proposed future colliders
for new heavy neutral ($Z'$) and charged ($W'$) gauge bosons are summarized.
Typical bounds achievable on $M_{Z',W'}$ at the TEVATRON,
DI-TEVATRON, LHC, 500 GeV NLC, and 1 TeV NLC are $\sim$1~TeV,
$\sim$2~TeV, $\sim$4~TeV, 1--3~TeV, and 2--6~TeV, respectively.
 For $M_{Z'} \sim$1 TeV
the LHC  will have the capability
 to  determine the magnitude of  normalized $Z'$
quark and lepton couplings to around $10-20\%$, while the NLC would allow for
determination of  the  couplings (including their signs)
with a factor of 2  larger error-bars, provided heavy flavor tagging
and   longitudinal polarization of the  electron beam is available.
\end{abstract}

\section{Introduction}

The existence of heavy neutral ($Z'$) and/or charged ($W'$) vector
bosons are a feature of many extensions of the standard model (SM).
They arise in extended gauge
theories including grand unified theories \cite{guts}, superstring
theories \cite{superstrings}, and
Left-Right symmetric models \cite{lrmodels} and in other models such as the
BESS model \cite{bess} and models of composite gauge
bosons \cite{composites}.

In this report we survey and compare
the discovery potential of the experiments that will be performed over the
next decade (TEVATRON, HERA, and LEP200) and future facilities that are being
planned and considered
for the period beyond (various TEVATRON upgrades, the LHC $pp$  collider,
the LSGNA 60~TeV $pp$ collider, the NLC $e^+e^-$ collider, and the
LEP-LHC $ep$ collider).  In addition to
the discovery reach of each of the experimental facilities we also address the
 auxiliary  question; that of the measurement of the properties of a
newly discovered gauge boson and therefore its identification in the
context of a specific model.
We therefore examine the diagnostic power of future colliders
for heavy gauge boson physics, in particular,  a (model independent)
determination of heavy gauge boson couplings to quarks and leptons.

In the next Section we give a brief description of the various models with new
gauge bosons that will be used in the study.
Given that  models without gauge invariance
will most  likely reveal themselves in ways other than the discovery of extra
gauge
bosons, we will restrict our analysis  to heavy gauge bosons that
arise in models with extended gauge symmetries.
In Section 3 we  summarize current constraints on heavy
gauge bosons.  In Section 4
we describe the signatures of extra gauge bosons at
future hadron and $e^+e^-$ colliders and the resulting discovery
reaches.  In Section 5 we address the  diagnostic power (identification
potential)  for heavy gauge boson physics at
the LHC 
and the NLC, respectively.  Conclusions are given in  Section 6.

There is a large and growing literature on this subject of which we
address only some aspects.
Our  primary goal is to explore the
 potential of experimental facilities for extra gauge bosons  physics
over the next decade
and beyond. The emphasis is on comparing the discovery limits
as well as the diagnostic power of the different facilities.
Other useful reviews which address related topics, some of them in
more detail, are:
Hewett and Rizzo \cite{h-r},  Hewett \cite{hewett}, Cveti\v c, del Aguila and
Langacker \cite{CAL},  and del Aguila \cite{aguila}.

\section{Models of Extra Gauge Bosons}

In this section we briefly describe some of the extended gauge theories
which have been studied in the literature.  While not totally
comprehensive, the properties are representative of models with extra gauge
bosons.

\subsection{Effective Rank-5 Models}

The largest set of extended gauge theories are those which are based
on GUTS \cite{h-r}.   Popular examples are the groups $SO(10)$ and $E_6$.
Generically, additional $Z$-bosons originating from $E_6$ grand unified
theories are conveniently labeled in terms of the chain
\begin{equation}
E_6 \to SO(10) \times U(1)_\psi \to SU(5)\times
U(1)_\chi \times U(1)_\psi \to SM \times U(1)_{\theta_{E_6}}
\end{equation}
where $U(1)_{\theta_{E_6}}$ remains unbroken at low energies.
Thus, the $Z'$ charges are given by linear combinations of the
$U(1)_\chi$ and $U(1)_\psi$ charges resulting in the $Z'$-fermion
couplings:
\begin{equation}
g_{Z^0} \left({g_{Z'}\over g_{Z^0}}\right)
(Q_\chi \cos \theta_{E_6} + Q_\psi \sin\theta_{E_6})
\end{equation}
where $\theta_{E_6}$ is a free parameter which lies in the range
$-90^\circ \leq \theta_{E_6} \leq 90^\circ$, $(g_{Z'}/g_{Z^0})^2
\leq {5\over 3}\sin^2\theta_w$ (here we assume the equality),
$Q_\psi = [1,1,1]/2\sqrt{6}$, and $Q_\chi  = [-1, 3, -5]/$ $2\sqrt{10}$
for $[(u,d,u^c,e^c), \; (d^c, \nu e^-), \;
(N^c)]$, the left-handed fermions in the {\bf 10}, $\overline{\bf 5}$,
and {\bf 1} of $SU(5)$ contained in the usual {\bf 16} of $SO(10)$.
Special cases of interest are model $\chi$ ($\theta_{E_6}=0^\circ$)
corresponding to the extra $Z'$ of $SO(10)$, model $\psi$
($\theta_{E_6}=90^\circ$) corresponding to the extra $Z'$ of $E_6$,
and model $\eta$ ($\theta_{E_6}=\arctan -\sqrt{5/3}$)
corresponding to the extra $Z'$ arising in some superstring
theories \cite{superstrings}.

\subsection{Left-Right Symmetric Model (LRM)}

$SO(10)$ GUTS lead to intermediate symmetries, for example;
\begin{eqnarray}
SO(10) & \to & SU(3)_C \times SU(2)_L \times U(1)_Y \times U(1)_\chi
\nonumber \\
& \to & SU(3)_C \times SU(2)_L \times SU(2)_R \times U(1)_{B-L}
\end{eqnarray}
The first chain leads to the additional $Z$-boson, $Z_\chi$, mentioned
above, while the second chain yields the left-right symmetric model (LRM)
which extends the standard model
gauge group to $SU(2)_L \times SU(2)_R \times U(1)$ \cite{lrmodels}
with a right-handed charged boson as well as an additional neutral
current.
The $Z'$-fermion coupling is given by
\begin{equation}
g_{Z^0}
{1\over {\sqrt{\kappa - (1+\kappa)x_W}}} [x_W T_{3L} + \kappa (1-x_W
) T_{3R} -x_W Q]
\end{equation}
with $0.55 \leq \kappa^2 \equiv (g_R/g_L)^2 \leq 1-2$  \cite{chang},
$T_{3L(R)}$ the isospin assignments of the fermions under
$SU(2)_{L(R)}$, $Q$ the fermion electric charge and
$x_W=\sin\theta_W$.  We assume $\kappa=1$ in our analysis which
corresponds to manifest left-right symmetric gauge interactions.  Note
that the $T_{3L}$ assignments are the same as in the standard model
while the values of $T_{3R}$ for $u_R, \; d_R, \; e_R, \; \nu_R =
{1\over 2}, \; -{1\over 2}, \; -{1\over 2}, \; {1\over 2}$ and are
zero for left-handed doublets.

\subsection{Alternative Left-Right Symmetric Model (ALRM)}

Another extended model based on the second intermediate group is the
alternative left-right model (ALRM) which
originates from $E_6$ GUT's and is also based on the electroweak
gauge group $SU(2)_L \times SU(2)_R \times U(1)$  \cite{ma}.  Here the
assignments for $T_{3L(R)}$ differ from those of the usual LRM for
$\nu_{L,R}$, $e_L$, and $d_R$ with
$T_{3L(R)}(\nu_L)= {1\over 2} (-{1\over 2})$,
$T_{3L(R)}(e_L)= -{1\over 2} (-{1\over 2})$,
and $T_{3L(R)}(d_R)=0$.  The LRM and ALRM have identical $u$-quark,
$e_R$, and $d_L$ couplings.  In this model the right handed
$W$-boson carries lepton number and has odd R-parity avoiding the
usual constraints on the mass of right handed $W$'s.

\subsection{``Sequential'' Standard Model (SSM)}

The ``sequential'' Standard Model (SSM) consists of a $Z'$ with the same
couplings as  the SM $Z^0$ boson
couplings. Although it is not a  gauge invariant model
it is often used
for purposes of comparison.

\subsection{Un-unified Standard Model (UNSM)}

The un-unified standard model (UNSM) \cite{georgi} is based on the gauge
group $SU(2)_l \times SU(2)_q \times U(1)_Y$, i.e., left-handed
leptons (quarks) transform as doublets under $SU(2)_l$ ($SU(2)_q$)
and singlets under  $SU(2)_q$ ($SU(2)_l$), and right-handed fields
are singlets under both groups.  The $Z'$-fermion coupling takes the form
\begin{equation}
g_{Z^0} c_w \left( { {T_{3q}\over {\tan\phi}} -\tan\phi T_{3l} }\right)
\end{equation}
where $T_{3q(l)}$ is the third component of $SU(2)_{q(l)}$-isospin,
$c_w=\cos\theta_w$, and $\phi$ is a mixing parameter which lies in
the range $0.22 \leq \sin\phi \leq 0.99$.  We take $\sin\phi=0.5$ in
our calculations.  The $Z'$ is purely left
handed in this model.

There are numerous other models predicting $Z'$'s in the literature
\cite{othermodels} but the subset described
above has properties
representative of the broad class of
models,  at least for the purposes of comparing discovery
limits at high energy colliders.

In all of the above models the $Z-Z'$ mass matrix takes the form
\begin{equation}
M^2 = \left( \begin{array}{cc}
	M_Z^2 & \gamma M_Z^2 \\
	\gamma M_Z^2 & M_{Z'}^2
	\end{array} \right)
\end{equation}
where $\gamma$ is determined within each model once the Higgs sector is
specified.  The physical eigenstates are then
\begin{eqnarray}
Z_1 =Z' \sin \phi + Z \cos\phi \nonumber \\
Z_2 =Z' \cos \phi - Z \sin\phi
\end{eqnarray}
where $Z_1$ is currently being  probed
at LEP and $\tan 2\phi
=2\gamma M_Z^2/(M_Z^2 -M_{Z'}^2)$.  LEP measurements
constrain   $|\phi| < 0.01$ \cite{langacker} which is smaller than could
be observed at high energy collider experiments.  Without loss of
generality we will therefore
ignore $Z-Z'$ mixing in the remainder of this review.

In this report we will also assume that the  branching ratios
include decays into only {\it three ordinary families}.

\section{Present Limits}

Before proceeding to future colliders it is useful to list
existing bounds as a benchmark against which to measure future
experiments.  Constraints can be placed on the
existence of $Z'$'s either indirectly from fits to high precision
electroweak data \cite{langacker}--\cite{pdg}  
 or from direct searches at operating collider facilities \cite{cdf}.

There have been a number of fits to precision data
\cite{langacker,ew}.  We list results\footnote{The analysis for current limits
on extra gauge bosons in the  un-unified standard model (UNSM) was done in Ref.
\cite{CST}.}
of Langacker \cite{langacker94} in Table \ref{cvetictab1}
which includes the most recent (1993) LEP data and is an update of
the global analysis of electro-weak data
described in Ref. \cite{langacker}.  Two sets of results are
presented; an unconstrained fit with no assumptions on the Higgs
sector, and a constrained fit with specific assumptions on the Higgs
sector and therefore the $Z^0-Z'$ mixing angle.

The highest mass limits come from direct searches by the CDF
experiment at the Tevatron \cite{cdf}.  The CDF limits are obtained by looking
for high invariant mass lepton pairs that would result from Drell-Yan
production of  $Z'$s and $W'$s \cite{LRR,pp} and their subsequent decay
to lepton pairs;  $p\bar{p}\to Z' \to \ell^+\ell^-$ and $p\bar{p}
\to W' \to \\
\ell^\pm \not\!\!p_T$.
The most recent CDF 95 \% confidence level results based on ${\cal
L}_{int}=19.6\hbox{pb}^{-1}$ are listed in Table \ref{cvetictab1}.
The next run
with $\sim 75$~pb$^{-1}$ will increase the discovery reach by
roughly 100 GeV.

The direct TEVATRON bounds on $M_{W'_{LR}}$ (for the main production
channel $pp  \rightarrow W'\rightarrow e\nu_e$) are in the 600 GeV  region,
while indirect constrained
[unconstrained] bounds are in the 1.4 TeV [300 GeV] region.

\begin{table}[hp]
\begin{center}
\begin{minipage}{9.3cm}
\let\normalsize=\captsize
\caption{
Current constraints on $M_{Z'}$ (in GeV) for typical models from direct
production at the TEVATRON (${\cal L}_{int}=19.6$ pb$^{-1}$), as well as
indirect limits from a global electro-weak analysis (95\%
C.L.) \protect \cite{langacker94}. }
\label{cvetictab1}
\vskip.5pc
\small
\setlength{\tabcolsep}{9pt}
\begin{tabular}{|l|ccc|} \hline
&direct&indirect & indirect \\ [-3pt]
& & (unconstrained)&(constrained)\\
\hline
$\chi$ & $425 $ & $330$ & $920$ \\
$\psi$ & 415  & 170 & 170 \\
$\eta$ & $440$ & $220$ & $610$ \\
$LR$ & $445$ & $390$ & $1360$\\
$SSM$&505&960&  \\ \hline
\end{tabular}
\end{minipage}
\end{center}
\end{table}
\setlength{\tabcolsep}{9pt}

\section{Discovery Limits of Extra Gauge Bosons}

In this section we present
the discovery reach for some of the models which exist in the
literature, and were described in Section 2.  Although far
from exhaustive,
these models  form a  representative set for the purposes of comparison.

Bounds on extra gauge bosons  attainable from
low energy neutral current precision experiments, measurements
at the TRISTAN, LEP and SLC $e^+e^-$ colliders,
as well as at the HERA $ep$ collider
have been surpassed by direct limits obtained
at the TEVATRON -$p\bar{p}$ collider or will be from planned TEVATRON
upgrades \cite{capstick,godfrey}. We will therefore
restrict the analysis to LEP200, proposed TEVATRON
upgrades, the LHC-$pp$ collider, the LSGNA 60~TeV $pp$ collider,
the NLC-high luminosity $e^+e^-$ collider,
and the LEP-LHC $ep$ collider \cite{capstick,godfrey,h-r92}.

\subsection{Hadron Colliders}

The signal for a $Z'$ at a hadron collider consists of
Drell-Yan production of lepton pairs \cite{guts,pp,capstick,godfrey,ehlq}
with high invariant mass via
$p \stackrel{(-)}{p} \to Z' \to l^+ l^-$.
The cross section for the production of on-shell $Z'$s is given by
\cite{capstick}:
\begin{equation}
{\hbox{d} \sigma (pp\rightarrow f\bar f) \over \hbox{d} y}
={x_A x_B \pi^2 \alpha_{em}^2 (g_{Z'}/g_{Z^0})^4 \over 9 M_{Z'}
\Gamma_{Z'} }\left({C_L^f}^2 +{C_R^f}^2 \right) \sum_q
\left({C_L^q}^2 +{C_R^q}^2 \right) G_q^+(x_A,x_B,Q^2)
\end{equation}
where
\begin{equation}
G_q^+(x_A,x_B,Q^2)=\sum_q \left[f_{q/A}(x_A) f_{\bar q/B}(x_B) +
f_{\bar q/A}(x_A) f_{q/B}(x_B) \right]
\end{equation}

The cross section for $Z'$ production at hadron colliders
is inversely proportional to the $Z'$  width.  If exotic
decay modes are kinematically allowed, the $Z'$ width will become larger and
more significantly
the branching ratios to conventional fermions smaller.
This is not important in $e^+e^-$ and $ep$ collisions since those processes
proceed via virtual $Z'$s in contrast to hadron colliders which
rely on the Drell-Yan production of real $Z'$s.
Having said this we will only consider the case
that no new decay modes are allowed.
The partial widths are given (at tree level) by
\begin{equation}
\Gamma_{Z' \rightarrow f \overline f } =
M_{Z'} g^2_{Z'} (C'^2_{f_L} +C'^2_{f_R})\big/24\pi
\end{equation}

\begin{figure}[ht]
\let\normalsize=\captsize
\begin{center}
\begin{minipage}{12.0cm}
\centerline{\psfig{file=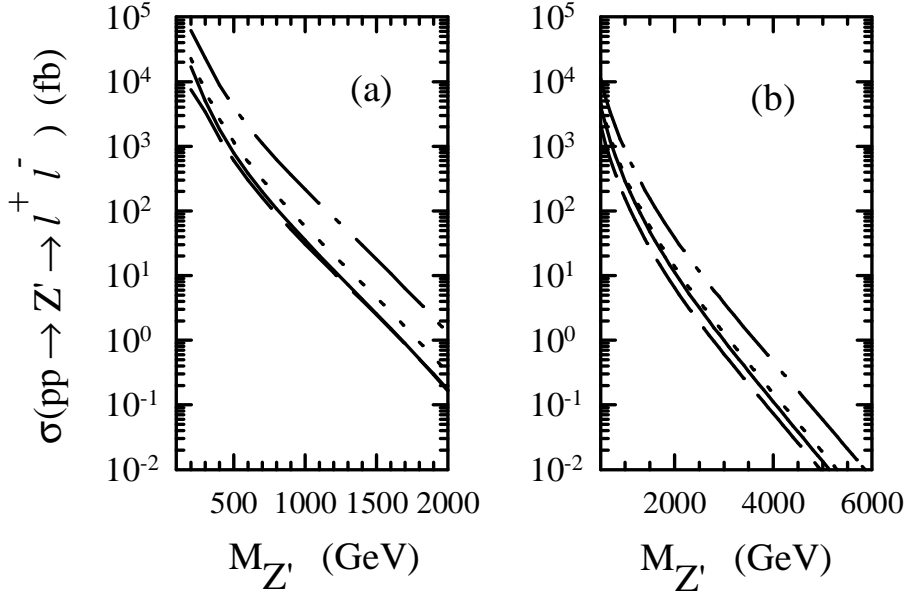,width=12.0cm}}
\caption[]{The cross section for the process $pp \to Z' \to
\ell^+\ell^-$ as a function of $M_{Z'}$
for (a) $p\bar{p}$ with $\sqrt{s}=4$~TeV and (b) $pp$
with $\sqrt{s}=14$~TeV.  In both cases the solid line is for $Z_\chi$,
the dashed line for $Z_\psi$, the dotted line for $Z_{LR}$ and the
dot-dashed line for $Z_{ALR}$.}
\label{cveticfig1}
\end{minipage}
\end{center}
\end{figure}

The cross section for $\sigma(pp \to Z')\cdot \hbox{BR}(Z'\to
\ell^+\ell^-)$ is shown in Fig.~\ref{cveticfig1}
as a function of $M_{Z'}$ for an upgraded
Tevatron ($p\bar{p}$) with $\sqrt{s}=4$~TeV and the LHC ($pp$) with
$\sqrt{s}=14$~TeV.   If we use 10 dilepton events
clustered at a particular invariant mass as the criteria for discovery of
a $Z'$ one can read off the discovery reach as the cross section times
integrated luminosity which gives this number of dilepton events.

We obtain the discovery limits for this process based on 10 events
in the $e^+e^- + \mu^+\mu^-$ channels using the EHLQ structure
functions \cite{ehlq} set 1, taking $\alpha=1/128.5$,
$\sin^2\theta_w=0.23$, and including a 1-loop $K$-factor in the $Z'$
production \cite{k-factor}. We include a $t$-quark of mass 174~GeV in the
$Z'$ decay
width, and 2-loop QCD radiative corrections and 1-loop QED radiative
corrections in calculating the $Z'$ width.  Using different
quark distribution functions results in a roughly 10\% variation in
the $Z'$ cross sections \cite{rizzo} with the subsequent change in
discovery limits.  We note that including
realistic detector efficiencies would lower these limits.

Lowering the number of events in the $e^+e^- + \mu^+\mu^-$ channels
to 6 raises the discovery
reach  about $10\%$ while lowering the
luminosity by a factor of ten  reduces the reach  by about a factor
of 3 \cite{capstick}.

In our calculations we assumed that the $Z'$ only decays into the
three conventional fermion families.  If other decay channels were
possible, such as to exotic fermions filling out larger fermion
representations or supersymmetric partners, the $Z'$ width would be
larger, lowering the discovery limits.  On the other hand, if decays
to exotic fermions were kinematically allowed, the
discovery of exotic fermions would be an important discovery in
itself;  the study of the corresponding decay modes would provide additional
information on the  nature of the extended gauge structure.

The discovery limits for various models at hadron colliders
are listed in  Table \ref{cvetictab2} and, for ease of comparison,
are shown in Fig.~\ref{cveticfig3} along with those of $e^+e^-$
colliders, both given at the end of this section.
These  bounds are relatively insensitive to specific models.  In addition,
since they are based on a distinct signal with little background
they are relatively robust limits.  For the case of the DI-TEVATRON
($\sqrt{s}=4$~TeV), the $p\bar{p}$ option has a 50\% higher discovery
reach than the $pp$ option
for a given luminosity indicating that valence quark contributions
to the Drell-Yan production process are still important at these
energies.

Discovery limits for $W'$ bosons obtained using the process
$pp \; (p\bar{p}) \to
W'^{\pm } \to \ell^\pm \nu_\ell$ and are based on 10 events in the
[ $(e^+\nu_e+ e^-\bar\nu_e)\ +\ (\mu^+\nu_\mu+ \mu^-\bar\nu_\mu)$]
channels.  They are given in Table \ref{cvetictab2}.
The $W'$ discovery limits are  larger than the comparable $Z'$
limits which is in part due
to the fact that in  most models with $W'$'s
[{\it e.g.}, the left-right symmetric model(s)] the $W'$  couplings
are comparable in magnitude with those of
the standard model $W$.
These results are based \cite{PLUS} on conservative assumptions  on the form of
the CKM
matrix, {\it i.e.}, the right-handed and left-handed quark-mixing matrix
elements have
an equal magnitude.  If these  assumptions  are relaxed \cite{PLUS, RX}, the
bounds could
change drastically.

\subsection{\protect\boldmath$e^+e^-$ Colliders}

At $e^+e^-$ colliders discovery limits are indirect, being inferred
from deviations from the standard model predictions for various
cross sections and asymmetries due to
interference between the $Z'$ propagator and the $\gamma$ and $Z^0$
propagators \cite{e+e-}.  This effect is similar to PEP/PETRA seeing the
standard model
$Z^0$ as deviations from the predictions of QED.
The basic process is $e^+e^- \to f\bar{f}$ where $f$
could be leptons $(e,\; \mu ,\; \tau)$ or quarks $(u, \; d, \; c,\;
s,\; b)$.
For $e^+e^-$ collider measurements all results are derived from the
differential cross section for a polarized $e^-$ and an unpolarized
$e^+$ \cite{capstick}:
\begin{equation}
{\hbox{d}\sigma (e^+e^-_L \rightarrow f\bar f )\over
\hbox{d}\cos \theta} = {\pi \alpha^2\over 4s}
\left\{ | C_{LL} |^2 (1+\cos\theta )^2
+| C_{LR}|^2 (1-\cos \theta )^2 \right\}
\end{equation}
where
\begin{eqnarray}
 C_{ij} = -Q_f + {C_i^e C_j^f \over \sin^2\theta_W\cos^2\theta_W}
&&\left({s \over {s-M_{Z^0}^2 + i\Gamma_{Z^0} M_{Z^0}}}\right)\\ \nonumber
&& + {(g_{Z'}/g_{Z^0})^2{C_i^e}'{C_j^f}'\over\sin^2\theta_W\cos^2\theta_W}
\left({s \over {s-M_{Z'}^2 + i\Gamma_{Z^{\prime}} M_{Z'}}}\right)
\end{eqnarray}

{}From these basic reactions
the following probes are used to search for the effects of $Z'$'s:
\begin{itemize}
\item The leptonic cross
section, $\sigma (e^+ e^- \to \mu^+ \mu^-)$.
\item The ratio of the hadronic to the QED point cross section,
$R^{had}= \sigma^{had}/\sigma_0$.
\item The leptonic forward-backward
asymmetry, $A^\ell_{FB}$,  and if $c$ and $b$
quark flavor tagging were sufficiently efficient, one could
measure  $A_{FB} (e^+e^- \rightarrow c\bar c)$ and
$A_{FB} (e^+e^- \rightarrow b\bar b)$. The forward-backward
asymmetries are given by:
\begin{equation}
A_{FB} = { {[\int^1_0 -\int_{-1}^0 ] d\cos\theta {d\sigma\over d\cos\theta}}
\over {[\int^1_0 +\int_{-1}^0 ] d\cos\theta {d\sigma\over d\cos\theta} } }
\end{equation}
\item The leptonic  longitudinal asymmetry,
$A^\ell_{LR}$, the hadronic longitudinal asymmetry,
$A^{had}_{LR}=A_{LR}(e^+e^-\to hadrons)$, and final state
polarization of $\tau$'s,
$A_{pol}^\tau$. These polarization asymmetries are defined by
\begin{equation}
A_{LR} = {\sigma(e^-_L) -\sigma (e^-_R) \over
\sigma(e^-_L) +\sigma (e^-_R)}
\end{equation}
where the cross sections are obtained by integrating eq. 4.4 over
$\cos \theta$.
(For electron polarization less than 100\% the asymmetry is given by
$A^P_{LR}= PA_{LR}^{P=1}$.)
Since the
cross section to hadrons is much larger than to leptons this results
in higher statistics and therefore more precise measurements.
\item
The polarized forward-backward asymmetry for
specific fermion flavors, $A^f_{FB}(pol)$  is obtained by considering the
forward-backward asymmetry for specific initial electron
polarizations.
\end{itemize}
In these expressions the indices $f=\ell, \;
q$, $\ell =(e,\mu,\tau)$, $q=(c, \; b)$,
and $had=$`sum over all hadrons' indicate the final state fermions.

\begin{figure}[htbp]
\let\normalsize=\captsize
\begin{center}
\begin{minipage}{12.0cm}
\centerline{\psfig{file=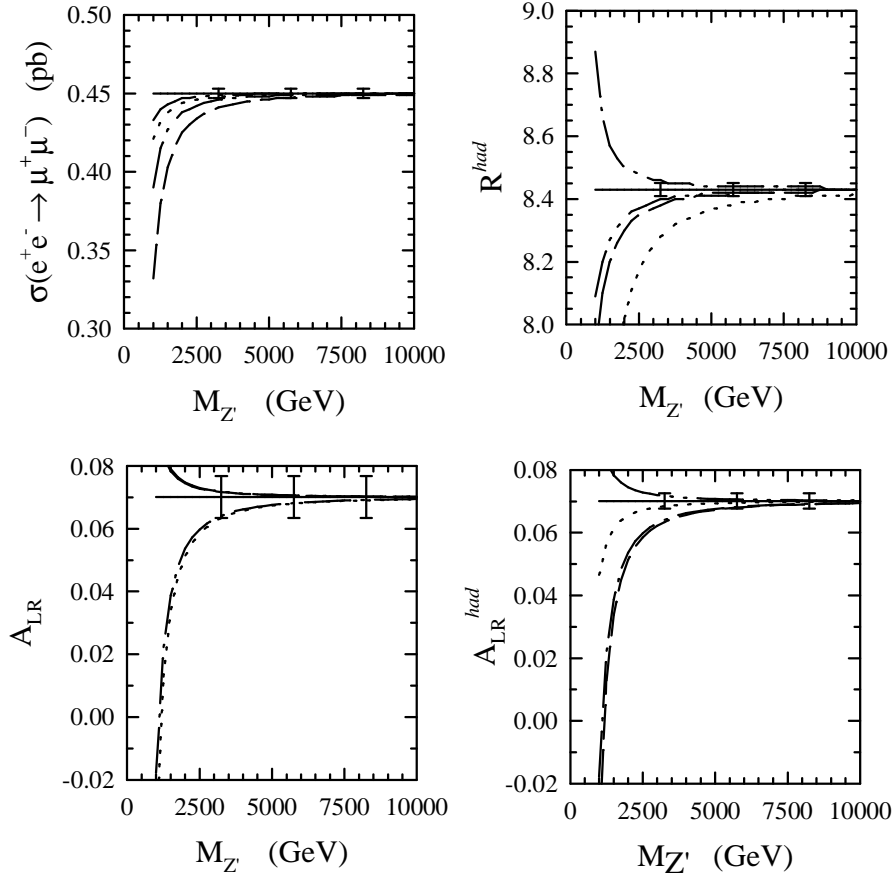,width=12.0cm}}
\caption[]{$\sigma(e^+e^- \to \mu^+\mu^-)$, $R_{had}$,
$A_{LR}$, and $A_{LR}^{had}$ for $\sqrt{s}=500$~GeV
as a function of $M_{Z'}$.  The error bars are statistical errors
based on 50~fb$^{-1}$ integrated luminosity.  In all cases the solid
line is the standard model prediction, the dot-dash line is for
model-$\chi$, the dot-dot-dash model is for model-LR, the dashed
line is for model-ALR and the dotted line is for model-SSM.}
\label{cveticfig2}
\end{minipage}
\end{center}
\end{figure}

For indirect limits, a 99\% C.L. corresponds to a $2\sigma$ effect of
one observable.
Since $2\sigma$ deviations are not uncommon one must be cautious
about how one obtains discovery limits for $Z'$'s.  One possibility
for obtaining believable limits is to raise the deviation required to
indicate the existence of a $Z'$.  A second possibility is to
combine several observables to obtain a $\chi^2$ figure of merit.
We follow the second approach here by including $\sigma^l$,
$R^{had}$, $A_{LR}$, and $A^{had}_{LR}$ to obtain the 99\% confidence
limits.  In Fig.~\ref{cveticfig2}
we show these observables for representative
models as a function of $M_{Z'}$.  1-$\sigma$ statistical errors are
shown for comparison with the deviations resulting from
the existence of a $Z'$.
The discovery limits obtained this way are given
in Table \ref{cvetictab2} and
shown in Fig.~\ref{cveticfig3} at the end of this
section.\footnote{Although it is far from
clear whether LEP200
will achieve any significant longitudinal polarization, $A_{LR}^{had}$
only  contributes significantly to the limit on $Z_\chi$ at LEP200 so
that our results are not in general sensitive to the inclusion of this
observable in the $\chi^2$ at LEP200.}

One sees that the discovery limits obtained at $e^+e^-$ colliders
are as high or higher than those that can be obtained at hadron
colliders.  However, the bounds obtained are more model dependent than
the  bounds obtained at hadron colliders.  For example, for model
$\psi$,
$C^\prime_L=\pm C^\prime_R$ so that either $C^\prime_V$ or $C^\prime_A=0$.
For $\sqrt{s}$ sufficiently far away from the $Z^0$ pole deviations are
dominated by $Z^0-Z'$ and $\gamma-Z'$ interference which is proportional
to $C_V^2 C_V^{\prime2} +2C_V C_A C_V^\prime C^\prime_A +C_A^2 C_A^{\prime2}$.
Since for the photon $C_A=0$ so when $C^\prime_V$ is also equal to 0,
deviations
from the standard model become small.

Because
the bounds obtained at $e^+e^-$ colliders are indirect, based on deviations
from the standard model in precision measurements, they are sensitive to
the experimental errors, both statistical and systematic.  For example,
reducing the LEP200 integrated luminosity from 500~pb$^{-1}$ to
250~pb$^{-1}$ reduces the discovery limits by about 15\% and reducing
the NLC integrated luminosity from 50~fb$^{-1}$ to 10~fb$^{-1}$
(200~fb$^{-1}$ to 50~fb$^{-1}$) for the 500~GeV (1~TeV) case reduces the
discovery limit by about 33\%.  Including a 5\% systematic error in
cross section measurements and a 2\% systematic errors in asymmetries
where systematic errors partially cancel \cite{barklow} can lower the
discovery limits significantly.  The most extreme change is
for the sequential standard model $Z'$ which decreases by a factor of 2
at LEP200 and a factor of 3 at the NLC.  Clearly, systematic errors will
have to be kept under control for high precision measurements.

 We did not include radiative corrections in our results.
In general, this is an acceptable procedure since we are looking for
small deviations from the standard model predictions and radiative
corrections to $Z'$ contributions will be a small correction to a small
effect.  However, QED bremsstrahlung corrections, in particular initial
state radiation, can give large contributions to the observables,
altering the statistics we assumed.  Since these are dependent on
details of the detector we have left them out, but note that they can
alter the numerical values.

\subsection{\protect\boldmath$ep$ Colliders}

A final type of collider that exists and is being considered for the
future are $ep$ colliders; HERA at DESY and LEP-LHC and LEP2-LHC at
CERN \cite{capstick87,ep}.  As in the $e^+e^-$ case
$Z'$ discovery is based on deviations from the
standard model for high precision measurements of various
observables. The diagrams are similar to those of the
hadron-collider and $e^+e^-$ cases except that the gauge bosons are
exchanged in the $t$-channel in contrast to $s$-channel production
in the former cases.  The differential
cross section for $ep$ collisions is given by:
\begin{eqnarray}
{d\sigma (e^-_Lp)\over dx\; dy}  = {2\pi \alpha ^2\over s\; x^2y^2}
\sum _q \Big\lbrace x\ f_q(x,Q^2) & & \left[ \vert b_{LL}\vert ^2
+\vert b_{LR}\vert ^2 (1-y)^2 \right] \\ \nonumber
& &  +x\ f_{\bar q}(x,Q^2) \left[ \vert b_{LR}\vert
^2+\vert b_{LL} \vert ^2(1-y)^2 \right] \Big\rbrace
\end{eqnarray}
where the sum runs over quark flavors. $f _q(x,Q^2)$ and $f_{\bar
q}(x,Q^2)$ are the quark and antiquark distribution functions, $Q^2=xys=-q^2$,
and $x$ and $y$ are the usual scaling variables, $x=Q^2/2p\cdot q$,
$y=p\cdot q /p\cdot k$.
The functions $b_{ij}$ are given by
\begin{equation}
b_{ij}= - Q_q + {C^e_iC^q_j \over \sin^2\theta_W \cos^2\theta_W }
 {Q^2\over Q^2+M^2_{Z^0}}
+\left( {g_{Z'}\over g_{Z^0}} \right) ^2
{ {C^e_i}' {C'^q_j}' \over \sin^2\theta_W \cos^2\theta_W}
{Q^2\over Q^2+M^2_{Z'}}
\end{equation}
where $Q_q$ denotes the quark electric charge and  $C_{L,R}$ and $C'_{L,R}$
are the left and right-handed $Z^0$ and $Z'$ charges.
For the $e^+_L p$ cross section
take $b_{LL}\to b_{RL}$ and $b_{LR}\to b_{RR}$
and to obtain the cross section for right-handed electrons and positrons make
the
substitution $L\leftrightarrow R$.

Given longitudinal polarization of incident $e^-$ or
$e^+$  there are eight different
measurements (of which four are independent);
the electron and positron cross sections $\sigma (e^-)$ and
$\sigma (e^+)$, and the six asymmetries $e^-_L - e^-_R$, $e^-_L - e^+_R$,
$e^-_R - e^+_L$, $e^+_L - e^+_R$, $e^-_R - e^+_R$, and $e^-_L - e^+_L$
where the asymmetry $\alpha - \beta$ is defined as
\begin{equation}
A_{\alpha \beta}
={{\sigma (\alpha) - \sigma (\beta)}
\over{\sigma (\alpha) + \sigma (\beta)}}.
\end{equation}

To obtain discovery limits based on deviations from the standard
model we follow the same approach used in obtaining limits at
$e^+e^-$ colliders.  We base our limits using a $\chi^2$ analysis of
the four observables; $A(e^-_L - e^-_R)$, $A(e^-_L - e^+_R)$,
$A(e^-_R - e^+_L)$, and $A(e^+_L - e^+_R)$.  Unlike the $e^+e^-$
analysis the deviations vary significantly for the different
observables \cite{capstick,capstick87}
and the $\chi^2$ tends to be dominated by only one or two of them.

We used the EHLQ structure functions \cite{ehlq} (set 2),
integrated over the $x$ and $y$ variables from 0.1 to 1.
The lower bound was so chosen because at small $x$ the
cross section is larger giving better statistics, but with smaller
deviations, while at large x the deviations are larger, but the statistics are
poorer.  Thus, we take $x_{min}=0.1$ as a reasonable compromise with
 adequate statistics which is not overwhelmed by the standard
model contributions.  More sophisticated event binning would likely
improve our bounds.

The results for the various $ep$ options are given in
Table \ref{cvetictab2}.
Despite the fact that for HERA
we used the overly optimistic integrated luminosity
of 600~pb$^{-1}$ per polarization the
discovery limits for HERA have already been surpassed by existing
Tevatron results.  Similarly, the discovery limits at the
LEP-LHC $ep$  collider
are not at all competitive with the LHC and are  at the verge of
being excluded by the TEVATRON.  The higher energy, but lower
luminosity LEP2-LHC option are essentially irrelevant.  Even if
LEP2-LHC could achieve integrated luminosities of 1~fb$^{-1}$ the
discovery limits would only yield a small improvement over the  bounds
achievable at   (the lower  energy) LEP-LHC option.

\begin{table}[hp]
\begin{center}
\let\normalsize=\captsize
\caption[]{%
Bounds on $M_{Z'}$ and $M_{W_{LR}^\prime}$ (in GeV)
for typical models achievable at
proposed hadron and $e^+e^-$ colliders.
The discovery limits  for $Z'$  [$W^{\prime +}+W^{\prime -}$]
at hadron colliders  are
for typical models  with  10 events in   $e^+e^-\ +\ \mu^+\mu^-$
[($e^+\nu_e +e^-\bar\nu_e\ $)  +\
($\mu^+\nu_\mu+\mu^-\bar\nu_\mu)$]
while those for
$e^+e^-$ colliders are 99\% C.L. obtained from a $\chi^2$ based on
$\sigma (e^+e^- \to \mu^+\mu^-)$, $R^{had}=\sigma (e^+e^- \to
hadrons)/\sigma_0$, $A_{LR}^{\mu^+\mu^-}$, and $A^{had}_{LR}$. }
\label{cvetictab2}
\renewcommand{\tabcolsep}{5.75pt}
\renewcommand{\arraystretch}{1.2}
\vskip1pc
\small
\begin{tabular}{|l|cc|ccccc|}
\hline
Collider &$\sqrt{s}$ [TeV] &${\cal L}_{int}\  [\hbox {fb}^{-1}]$& $\chi$
& $\psi$ & $\eta$ & $LR$ & $W'_{LR}$ \\
\hline
TEVATRON ($p\bar{p})$ &
 \91.8& \9\91&  \9775 & \9775 & \9795 & \9825 & \9920 \\
TEVATRON$^\prime$ ($p\bar{p})$ &
 \92\9\ & \910&  1040 & 1050 & 1070 & 1100 & 1180 \\
DI-TEVATRON ($p\bar{p})$ &
 \94\9\ & \920&  1930 & 1940 & 1990 & 2040 & 2225 \\
LHC ($pp)$ &
  10\9\ & \940&  3040 & 2910 & 2980 & 3150 &  \\
LHC ($pp)$ &
  14\9\ & 100 &  4380 & 4190 & 4290 & 4530 & 5310 \\
\hline
LEP200 ($e^+e^-$) &
  0.2& \9\90.5& \9695 & \9269 & \9431 & \9493 &  \\
NLC ($e^+e^-$) &
  0.5&\9  50\9\ &  3340 & \9978 & 1990 & 2560 &  \\
NLC-A ($e^+e^-$) &
  1.0 & 200\9\ & 6670 &  1940 & 3980 & 5090 &  \\
NLC-B ($e^+e^-$) &
  1.5 & 200\9\ & 8220 &  2550 & 4970 & 6240 &  \\
NLC-C ($e^+e^-$) &
  2.0 & 200\9\ & 9560 &  3150 & 5830 & 7210 &  \\
\hline
HERA ($ep$) &
  0.314 & 0.6 & 235 & 125 & 215 & \9495 &  \\
LEP$\times$LHC ($ep$) &
  1.183 & 1.0 & 375 & --- & 435 &  1060 &  \\
LEP2$\times$LHC ($ep$) &
  1.670 & 0.1 & --- & --- & --- & \9615 &  \\
\hline
\end{tabular}
\end{center}
\end{table}

\begin{figure}[htbp]
\let\normalsize=\captsize
\begin{center}
\centerline{\psfig{file=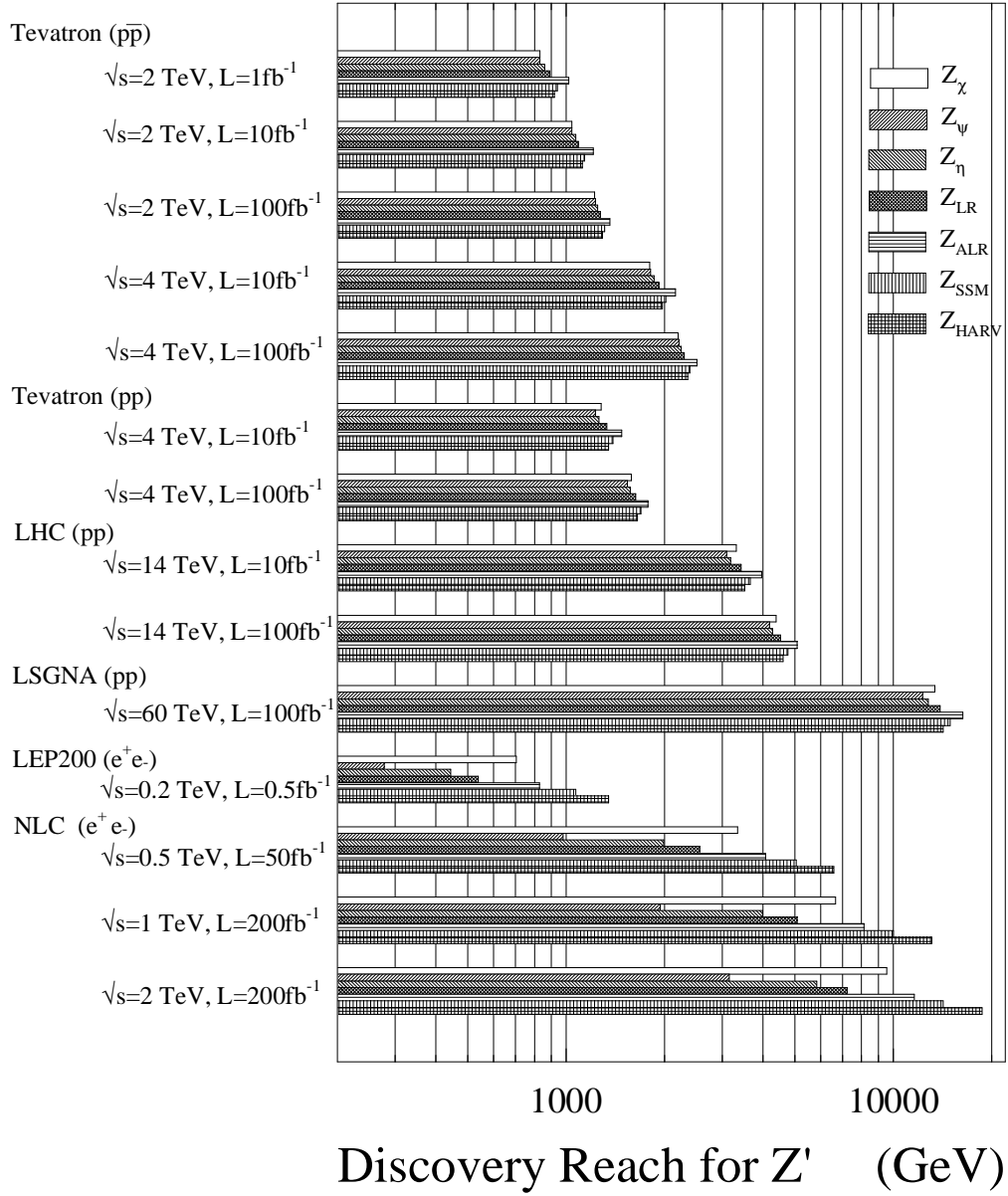,width=14.0cm}}
\begin{minipage}{14.0cm}
\caption[]{Discovery limits for extra neutral gauge bosons ($Z'$)
for the models described in the text.  The discovery limits at
hadron colliders are based
on 10 events in the $e^+e^-\ +\ \mu^+\mu^-$ channels while those for
$e^+e^-$ colliders are 99\% C.L. obtained from a $\chi^2$ based on
$\sigma (e^+e^- \to \mu^+\mu^-)$, $R^{had}=\sigma (e^+e^- \to
hadrons)/\sigma_0$, $A_{LR}^{\mu^+\mu^-}$, and $A^{had}_{LR}$.
The integrated luminosities are based on a $10^7$~sec year of running.}
\label{cveticfig3}
\end{minipage}
\end{center}
\end{figure}

\clearpage

\section{New Gauge Boson Diagnostics at Future Colliders (LHC \& NLC)}

An immediate need after the discovery of a new gauge boson would be to
determine its origin by measuring  its properties \cite{ACL}:

\begin{enumerate}
\item $Z'$ couplings to quarks and leptons.

\item The nature of the symmetry breaking sector.

\item $Z'$ couplings to exotics.
\end{enumerate}

A test of the  symmetry breaking structures are the decays $Z'\to
W^+W^-$ \cite{ZWIRNER,CKL},  which are suppressed by $Z-Z'$ mixing but
still  have a sizable rate due to the
enhancement of  the longitudinal components of the $W$ bosons.
However, they suffer from serious QCD backgrounds \cite{GUNION,AGUILA}.
In  theories with  charged gauge bosons, {\it e.g.}, left-right (LR)
symmetric models, the ratio $M_{Z'}/M_{W'}$ plays an  analogous
role to
the $M_Z/M_W$  ratio (related
to the  $\rho$ parameter)   in the standard model.
 This ratio
therefore  yields indirect  information on the nature of the
Higgs sector \cite{CLO}.

The study of $Z'$ decays into exotic
particles  also  yields  useful information.
In particular, $W'\to \ell N$ and $Z'\to NN$ and subsequent  decays of
heavy right-handed neutrinos  $N$ turn out
to be useful probes for distinguishing
the left-right models from those with only an additional $U(1)$ \cite{CKL,DL}.

 In the following we will concentrate on the diagnostic study of new gauge
boson couplings  to quarks and leptons.  In
 Subsection 5.1 we address the new gauge boson diagnostics at
 the LHC.  In Subsection 5.2 we present an analogous
 analysis at the NLC. The comparison of the diagnostic power of
 the two types of machines is given in Subsection  5.3.

\subsection{New Gauge Boson Diagnostics at the LHC }

In the main production channels,  $pp \to Z'\to \ell^+
\ell^-$ ($\ell=e,\mu$), one would be able to measure immediately
the mass $M_{Z'}$, the width $\Gamma_{tot}$ and the total cross section
 $\sigma_{\ell \ell}$. However, $\sigma_{\ell\ell}=\sigma (pp\to Z')
 B$  is {\it not}   a useful diagnostic
probe for the $Z'$ gauge couplings to quarks and leptons.
While $\sigma(pp\to Z')$,   the total  production cross section, can be
calculated to within  a few percent for
given $Z'$ couplings, the branching ratio into leptons,
$B\equiv\Gamma(Z'\to\ell^+\ell^-)/\Gamma_{tot}$,
is model dependent; it depends on the contribution of exotic
fermions and supersymmetric partners to the
$Z'$ width, and thus it  cannot be
useful as a diagnostic test for the $Z'$ gauge couplings.  However, it
would be a useful indirect probe for the existence of the exotic
fermions or superpartners.
On the other hand, from measurements of the total width
$\Gamma_{tot}$,  and $\sigma_{\ell \ell}$ one obtains
$\sigma \Gamma(Z'\rightarrow \ell^+\ell^-)\equiv\sigma B\Gamma_{tot}$, which
 probes the absolute magnitude of the gauge couplings.

In the following we
will  address signals  which probe {\it  relative strengths} of $Z'$ gauge
couplings. The  forward-backward asymmetry \cite{LRR} in
the main  production channel  $pp\to Z'\to
\ell^+\ell^- \ (\ell=e$ or $\mu$).\footnote{See also Ref. \cite{AV}. }\
was the first recognized  probe for the  gauge couplings
 at future hadron  colliders.
Since then a number of new,  complementary  probes were
proposed \cite{CKL,CLII}--\cite{RM}.

The nature of such probes can be classified according to the
type of  channel in which they can be measured:
\vspace{6pt}

\begin{description}
\item[(Ia)] {\it The main  production channels}:

  \begin{description}
  \item[(A)] Forward-backward asymmetry \cite{LRR},

  \item[(B)] Ratio of cross-sections in different rapidity bins \cite{ACL},

  \item[(C)] Corresponding asymmetries
    if proton polarization  were available \cite{FT}.
  \end{description}

\item[(Ib)] {\it Other two-fermion final state channels}:

  \begin{description}
  \item[(D)] Measurements of the $\tau$ polarization
    in the $pp\to Z'\to  \tau^+\tau^-$ channel \cite{AAC},

  \item[(E)] Measurements of the cross section in the $pp\to Z'\to
    jet\  jet$  channel \cite{RM,MOH}.
  \end{description}

\item[(II)] {\it  The four-fermion final state  channels}:

  \begin{description}
  \item[(F)] Rare decays $Z'\to W\ell\nu_{\ell}$ \cite{RI,CLII},

  \item[(G)] Associated productions $pp\rightarrow Z' V$
    with $V=(Z,W)$ \cite{CLIV}  and $V=\gamma$ \cite{RII}.
  \end{description}
\end{description}

Probes under  (Ia) constitute   distributions, {\it i.e.},
``refinements'', in the main production channels.
The  forward-backward asymmetry  (A)  is the cleanest one, probing
a particular combination of  quark and lepton couplings. On the other hand,
the rapidity ratio (B) \cite{ACL}  was recognized as
a useful complementary probe separating the $Z'$ couplings to
the $u$ and $d$ quarks due to the  harder valence $u$-quark distribution
in the proton relative to the $d$-quark.
Probes (C) are  useful  ones if
 proton polarization were available at future hadron colliders. In addition,
better knowledge of  the polarized  spin distribution functions for quarks is
needed.

 For probes in  other than  the main production channels (Ib)
the   background can be large. For  (E) recent studies indicate
\cite{PO,RM,MOH}\
that  the large  QCD background may
be overcome with
appropriate kinematic cuts, excellent
dijet mass resolution, and  detailed knowledge of the QCD backgrounds.
(D)  provides another interesting possibility to address the $Z'$  lepton
couplings, while (E) is the only  probe available for the left-handed quark
coupling \cite{CLIV}.

Probes in the  four-fermion final state  channels (II)
have suppressed rates  compared  to the two-fermion channels (Ia) and (Ib).
In these cases one hopes to  have enough statistics, and no attempt to study
distributions  seems to be possible.

Rare decays   $Z'\rightarrow f_1\bar f_2 V$,
with  ordinary  gauge bosons $V=(Z,W)$ emitted
by Brems--strahlung from one of the
fermionic  ($f_{1,2}$) legs turn out to have sizable statistics
\cite{CLII}, which
is  due to a  double  logarithmic
enhancement \cite{CLII}, closely related
to collinear and infrared singularities of gauge theories. They
 were studied in detail in Refs. \cite{CLII,PACO,HR}.
A background study \cite{CLII,CLIV} of
such  decays  revealed that the only useful  mode\footnote{$Z'\to Z
\ell^+\ell^-$
does not significantly discriminate between models.}  without
large standard model and QCD backgrounds
is (F): $Z'\to W\ell \nu_{\ell}$  and $W\to hadrons$,
with the imposed  cut $m_{T\ell\nu_\ell}>90$ GeV
on the transverse mass of the
$\ell \nu_\ell$.
(This assumes that there is a sufficiently high efficiency for the
reconstruction of $W\to hadrons$ in events tagged by an energetic lepton.
Further study of the QCD background and the jet reconstruction for such
processes is needed.)
 The same mode   with
$W\to \ell\nu_\ell$ may also   be detectable \cite{PACO}  if appropriate
cuts are applied.\footnote{ A   possibility
of gaining useful information from
$Z'\to Z\nu_{\ell}\bar\nu_{\ell}$ \cite{CLII,CLIV,HR}\
 was also addressed  in Ref. \cite{HRIII}.}
These modes probe a left-handed leptonic coupling.

 Associated productions  (F) turn out to be relatively clean
signals \cite{CLIV}  with slightly smaller statistics than rare decays.
They probe  a particular combination of the gauge couplings to quarks
and are thus complementary  to  rare decays.

 At  the LHC the above signals are feasible
 diagnostic probes  for $M_{Z'}\lsim 1-2$ TeV.
For diagnostic study of $Z'$ couplings large
luminosity is  important.
For higher $Z'$ masses the number of events drops rapidly. For
$M_{Z'}\simeq 2$ TeV, the statistical errors on forward-backward asymmetry
(A), the
rapidity ratio  (B), and rare decays (F)
 increase by a factor of  4, while those on associated productions (G)
 increase by a factor of 3.  A
reasonable discrimination between models and determination of
the couplings may still be  possible, primarily from the forward-backward
asymmetry and the rapidity ratio. However, for
$M_{Z'}\simeq 3$ TeV the statistical errors on the first three quantities
are larger by a factor of 13 than for 1 TeV, and there are not enough
events expected for
the associated production to allow a meaningful measurement.
 For $M_{Z'}\le 3$ TeV, there is therefore little ability
to discriminate  between models.

\subsubsection{Determination of Gauge Couplings  at the LHC}

We next examine how well the various $Z'$
couplings could be extracted from  the above probes. We will
mainly concentrate on probes (A), (B), (F) and (G), which seem to be
the  most feasible.
 For definiteness, we consider the statistical uncertainties for
  $M_{Z'}=$1 TeV at the LHC.
Eventually, the uncertainties associated with the detector acceptances
and  systematic errors will have to be taken into account.

In the following we assume family universality,
 neglect $Z-Z'$ mixing  and  assume $[Q',T_i]=0$,
which holds for $SU_2 \times U_1 \times U_1'$ and LR models.
 Here,  $Q'$ is the $Z'$ charge  and $T_i$ are the $SU_{2L}$
generators.\footnote{ For conventions
in the neutral current interactions see Ref. \cite{LL}.}

The relevant quantities \cite{CLIV,ACL}\  to
 distinguish different theories are
the charges, $\hat{g}^u_{L2}=\hat{g}^d_{L2}\equiv\hat{g}^q_{L2}$,
$\hat{g}^u_{R2}$, $\hat{g}^d_{R2}$, $\hat{g}^\nu_{L2}=\hat{g}^e_{L2}
\equiv\hat{g}^\ell_{L2}$, and $\hat{g}^\ell_{R2}$, and the gauge
coupling strength $g_2$.
The signs of the charges will be hard to determine at hadron
colliders. Thus the following
 four ``normalized''
observables can be probed \cite{CLIV}:
\def\denom{{(\hat{g}^\ell_{L2})^2+(\hat{g}^\ell_{R2})^2}}
\begin{equation}
\gamma_L^\ell\equiv{{(\hat{g}^\ell_{L2})^2}\over\denom},\ \
\gamma_L^q\equiv{{(\hat{g}^q_{L2})^2}\over\denom},\ \
\tilde{U}\equiv{{(\hat{g}^u_{R2})^2}\over {(\hat{g}^q_{L2})^2}},\ \
\tilde{D}\equiv{{(\hat{g}^d_{R2})^2}\over {(\hat{g}^q_{L2})^2}}.
\label{tild}
\end{equation}
The values of  $\gamma_L^\ell$, $\gamma_L^q$,
 $\tilde{U}$, and $\tilde{D}$  for the above models  are listed in
Table \ref{tlhc}.

\begin{table}[htbp]
\begin{center}
\begin{minipage}{11cm}
\let\normalsize=\captsize
\caption{ Values of the ``normalized''  couplings (1)
for the typical   models  and the statistical
error-bars as determined from probes at the LHC ($\protect\sqrt s = 14$ TeV,
 ${\cal L}_{int}=100\ \hbox{fb}^{-1}$). $M_{Z'}=1$ TeV.}
\label{tlhc}
\vskip.5pc
\small
\renewcommand{\arraystretch}{1.2}
\begin{tabular}{|r|cccc|} \hline
&$\chi$&$\psi$&$\eta$&$LR$\\ \hline
$\gamma^\ell_L$&$0.9\pm 0.016$&$0.5\pm 0.02$&$0.2\pm 0.012$&
$0.36\pm 0.007$\\
$\gamma^q_L$&0.1&0.5&0.8&0.04\\
$\tilde{U}$&$1\pm 0.16$&$1\pm 0.14$&$1\pm 0.08$&$37\pm 6.6$\\
$\tilde{D}$ & $9\pm 0.57$ & $1\pm 0.22$ & $0.25\pm 0.16$ & $65\pm 11$ \\
\hline
\end{tabular}
\end{minipage}
\end{center}
\end{table}

The forward-backward asymmetry  (A) is defined as:
\begin{equation}  A_{FB}= {{\left[\int_{0}^{y_{max}}-\int_{-y_{max}}^0\right]
[F(y)-B(y)]dy}\over
 {\int_{-y_{max}}^{y_{max}} [F(y)+B(y)]dy}}\ ,
\end{equation}
  while the
 rapidity ratio (B) is defined as \cite{ACL}:
\begin{equation}r_{y_1}= {{\int_{-y_1}^{y_1}[F(y)+B(y)]dy}
\over{(\int_{-y_{max}}^{-y_{1}}+\int_{y_1}^{y_{max}}) [F(y)+B(y)]dy}}\ .
\end{equation}
Here $F(y)\pm B(y)=[\int_0^1\pm\int_{-1}^0] \,d\cos\theta (d^2\sigma/
dy\, d\cos\theta)$, where  $y$  is the $Z'$ rapidity and $\theta$ is the
$\ell^-$ angle in the $Z'$ rest
frame. The rapidity range is  from $\{ -y_{max}, y_{max}\}$.
 $y_1$ is chosen in a range $0<y_1<y_{max}$ so that the
number of events in the two bins are comparable.
 At the LHC ($y_{max}\simeq 2.8$) for  $M_{Z'}\simeq 1$ TeV,  and $y_1=1$
 turns out to be an appropriate choice.
For rare decays (F)  one defines \cite{CLII}:
\begin{equation}
r_{\ell\nu W} \equiv {{B(Z'\rightarrow W\ell\nu_\ell)}
\over {B(Z'\rightarrow\ell^+\ell^-)}}\ ,
\end{equation}
 in which one sums
over $\ell=e,\mu$ and  over $W^+$, $W^-$.
For the associated productions (G)
 one   defines \cite{CLIV}
 the ratios:
\begin{equation}
R_{Z'V}={{\sigma
(pp\to Z'V)B(Z'\to \ell^+\ell^-)}\over{
\sigma (pp\to Z')B(Z'\to \ell^+\ell^-)}}\ ,
\end{equation}
 with $V=(Z, W)$\   decaying into leptons   and  quarks, and $V=\gamma$\
with an imposed $p_{T\gamma}\geq 50$ GeV cut.
 $\ell$ includes both
$e$ and $\mu$.

Statistical   errors  and explicit dependence of the above probes on
 the couplings (\ref{tild})
for $M_{Z'}=1$ TeV
  are given in Table \ref{cvetictab4}.
The EHLQ distribution functions \cite{ehlq},
 set 1, are used. One   also  assumes that the $Z'$ only
decays into  the ordinary three families of quarks and leptons.
Realistic fits, which include updated structure functions, kinematic cuts,
and detector  acceptances are expected to give larger uncertainties for
the couplings.\footnote{Table \ref{tlhc} updates
 the results  of  Ref.~\cite{ACL} where
the uncertainties are given for  $\sqrt{s}=16$ TeV. In addition, in
Table \ref{cvetictab4}
more optimistic assumptions on the branching ratios are  used.}

\begin{table}[htbp]
\begin{center}
\let\normalsize=\captsize
\caption{a) displays  the dependence of the probes on the couplings
[defined in (5.1)] at the
 at the LHC  ($\protect\sqrt s = 14$ TeV,
 ${\cal L}_{int}=100\ \hbox{fb}^{-1}$) for  $M_{Z'}=1$ TeV.  The
numerical values (with statistical errors) of the  corresponding probes for
typical models are given in  b).}
\label{cvetictab4}
\vskip.5pc
\begin{minipage}{6.5cm}
\leftline{a)}
\renewcommand{\arraystretch}{1.7}
\begin{tabular}{|l|c|}
\hline
$A_{FB}$ & ${{.387(2\gamma_L^l-1)\times(1-.753\tilde U-.247\tilde D)}
\over{1+.684\tilde U+.316\tilde D}}$\\
$r_{y_1}$ & $1.796{{1+.652\tilde U+.348\tilde D}\over{
1+.736\tilde U+.264\tilde D}}$\\
$A_{FB_{y_1}}$ & $ .726{{1-.731\tilde U-.269\tilde D}\over{
1-.769\tilde U-.231\tilde D}} $\\
$B_{qq}$ & $\gamma_L^l(2+\tilde U+\tilde D)  $\\
$r_{l\nu W}$ & $0.067\gamma_L^l
$\\
$R_{Z'Z}$ & ${{ 10^{-3}(7.55+.924\tilde U+0.098\tilde d)}\over{
1+.684\tilde U+.316\tilde D}}
$\\
$R_{Z'W}$ & $ {{24.53\times 10^{-3}}\over{
1+.684\tilde U+.316\tilde D}}$\\
$R_{Z'\gamma}$ & ${{ 5.38\times 10^{-3}(1+.896\tilde U+.104\tilde D)}\over{
1+.684\tilde U+.316\tilde D}}$\\
\hline
\end{tabular}
\end{minipage}
\renewcommand{\arraystretch}{1.3}
\vskip1pc
\begin{minipage}{14.5cm}
\leftline{b)}
\begin{tabular}{|l|cccc|}\hline
$$&$\chi$&$\psi$&$\eta$&$LR$\\ \hline
$A_{FB}$ &$-.1346\pm.0063$&$0\pm0.0087$&$-.0244\pm.0080$&$.1025\pm.0059$\\
$r_{y_1}$ &$2.091\pm.020$&$1.796\pm.023$&$1.732\pm.020$&$1.891\pm.017$ \\
$A_{FB_{y_1}}$ &$.85\pm.08$&$$&$.85\pm.58$&$.74\pm.09$\\
%
%
$r_{l\nu W}$&$.059\pm.0013$&$.033\pm.0014$&$.013\pm.0008$&$.023\pm.0008$  \\
$R_{Z'Z}$ &$2.08\pm.21$&$4.24\pm.4$&$4.88\pm.4$&$1.03\pm.13$ \\
$R_{Z'W}$ &$5.41\pm.33$&$12.07\pm.68$&$13.99\pm.67$&$.538\pm.097$ \\
$R_{Z'\gamma}$&$3.28\pm.26$&$5.23\pm.45$&$6.02\pm.44$&$4.9\pm.29$ \\
\hline
\end{tabular}
\end{minipage}
\end{center}
\end{table}

The error bars  turn out to be sufficiently small
 to distinguish  between models.
 The six quantities  $A_{FB}$,
$r_{y_1}$, $r_{\ell\nu W}$, and $R_{Z'V}$ with $(V=Z,W,\gamma)$
yield significant information on  three
($\gamma_L^\ell$, $\tilde U$ and $\tilde
D$) out of four normalized gauge couplings
 of ordinary fermions to the $Z'$.
The fourth normalized coupling    $\gamma_L^q$
could be determined \cite{CLIV}\ by a measurement of the branching ratio
$B(Z'\rightarrow q{\bar q})$. It turns out \cite{RM}, however, that for
$M_{Z'}\ge 1$ TeV  and the typical models  (except the SSM), the $Z'$ gauge
couplings are too small to  allow for determination of $\gamma^q_L$ with
sufficient  precision at the LHC.

\begin{figure}[htbp]
\let\normalize=\captsize
\centering{%
\centerline{\psfig{file=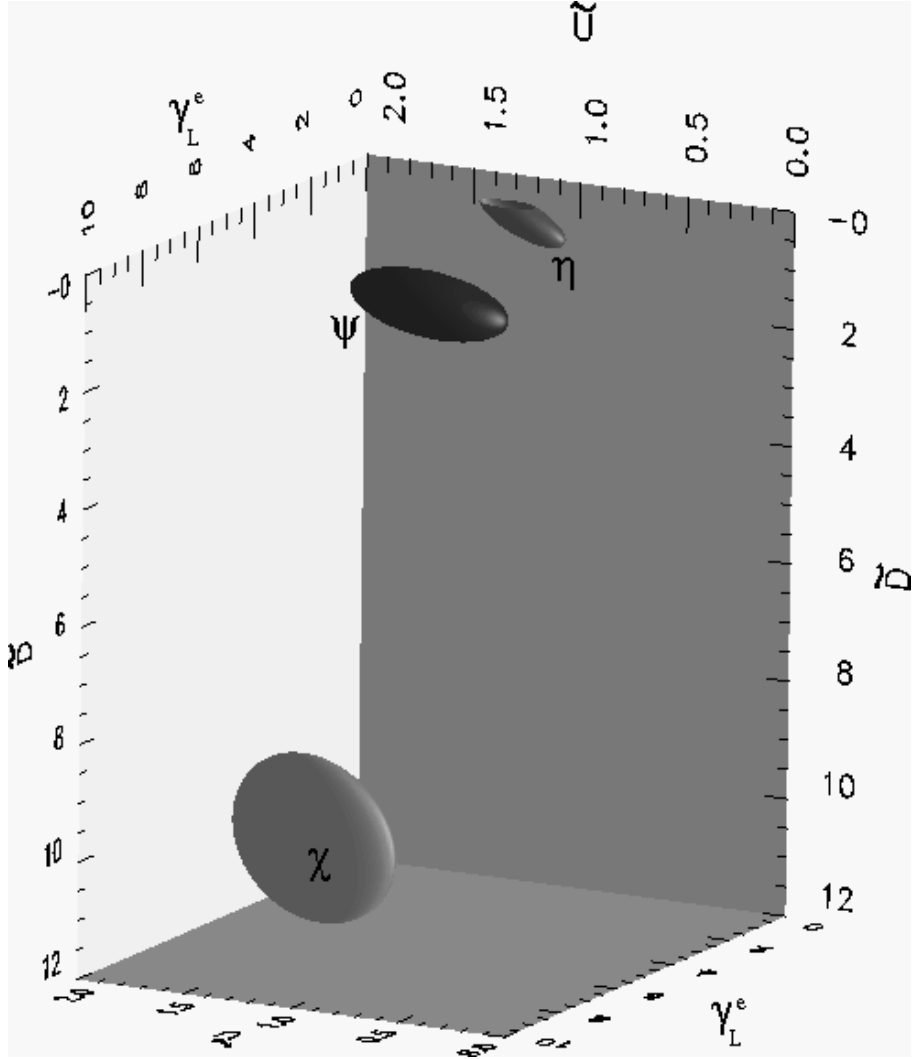,width=12.0cm}}
}
\caption[]{ 90\%  confidence level ($\Delta \chi^2=6.3$)
contours   for the $\chi$, $\psi$ and
$\eta$ models are plotted for $\tilde U$, versus  $\tilde D$, versus
$\gamma_L^\ell$. The input data are for $M_{Z'}=1$ TeV  at the LHC
($\protect\sqrt
s = 14$ TeV and  ${\cal L}_{int}=100\ \hbox{fb}^{-1}$)
and include statistical errors only.}
\label{cveticfig4}
\end{figure}

To study the precision to which these couplings could be determined,
a  combined $\chi^2$ analysis of these observables has been
performed, updating the earlier analysis of Ref. \cite{ACL}.
Only the statistical uncertainties have been included and
correlations between the observables have been ignored.
The results are  given in  Table \ref{tlhc}.
In particular, $\gamma_L^\ell$ can be determined very well (between
2\% and 8\% for the $\chi$, $\psi$, and $\eta$ models), primarily due to
the small statistical error for the rare decay mode $Z'\to W\ell\nu_\ell$.
On the other hand the  quark couplings have larger uncertainties,
typically 20\% .
In  Fig.~\ref{cveticfig4},
90\% confidence level  ($\Delta \chi^2=6.3$) contours\footnote{
The 90\% confidence level  contours for  projections
on the more familiar  two-dimensional
parameter subspaces correspond to $\Delta \chi^2=4.6$.}
are given in a three-dimensional plot for $\tilde U$ versus $\tilde D$
versus $\gamma_L^\ell$ for
the  $\eta$, $\psi$ and $\chi$ models. The $LR$ model
has  $\tilde U$ and $\tilde D$ in a
different region of the parameter space  (see
Table \ref{tlhc}).  From Fig.~\ref{cveticfig4} it is clear
that  one can distinguish well between  different  models.

%

The LHC can also address $W'$ diagnostics for $M_{W'}\lsim 1-2$
TeV. In particular,
indirect information on the Higgs sector is possible from $W'\rightarrow W Z$
decays.  The ratio $M_{W'}/M_{Z'}$  (an analog of the $\rho$ parameter
in the standard model) would also  yield  information on the nature of the
Higgs
sector. $W'$  decays into right-handed neutrinos (and other exotics)
would yield  additional useful information  on the $W'$.

$W'$ gauge couplings to ordinary quarks and
leptons is possible  by studying the forward-backward
 asymmetry and the rapidity ratios in the
 main production channel $pp\to W'\to \ell\nu_{\ell}$ ($\ell=e,\mu$)
as well as  associated productions ({\it e.g.},
$pp\to W'Z$)  and rare decays
({\it e.g.}, $W'\to Z\ell^+\ell^-$) in the corresponding four-fermion final
state channels.\footnote{See for example  Ref. \cite{CLL}.}
While the forward-backward asymmetry
in the main production channels $pp\to W'\to \ell\nu_\ell$  ($\ell=e,\mu $)
probes some  combination of gauge couplings, it does not distinguish $\hat
g_{L2}$
from $\hat g_{R2}$ couplings. On the
other hand,  rare decays
$W^{\prime\pm}\to W\ell^+\ell^-$ and associated productions
$pp\to W^{\prime\pm}W^\mp $ are strongly
suppressed \cite{CLII,CKL,CLIV}
 if $W'$ has only $\hat g_{R2}$ couplings, as in the LR symmetric models.
In models where $W'$
has $\hat g_{L2}$ couplings, {\it e.g.},
the so-called ununified standard model (UNSM) \cite{georgi}, the
corresponding rates are, however, not suppressed; primarily due to the
larger gauge couplings of the $W'$, the
corresponding  rates  allow for determination of $\hat g_{L2}$
couplings for $M_{W'}$
up to around 3 TeV. Note, however, that  for these processes it is difficult to
disentangle the contributions
from the direct coupling of the $W'$ to ordinary fermions and the
non-Abelian coupling of the $W'$ to  ordinary gauge bosons.

For $M_{Z'}\simeq 2$ TeV a reasonable
discrimination between models and determination of
the couplings may still be  possible, primarily from the forward-backward
asymmetry and the rapidity ratio. However, for
$M_{Z'}\simeq 3$  TeV there is little ability
to discriminate  between models.

\subsection{New Gauge Boson Diagnostics at the NLC}

If a $Z'$ were produced on shell at the NLC it would be relatively
straightforward to determine its properties.  On the other hand,
if it is far off-shell (a more likely
possibility) its
properties could be deduced \cite{DLRSV,HREE} through interference effects of
the $Z'$
propagator with the $\gamma$ and $Z$ propagator.
In this case the various observables described in the previous section to
deduce
the existence of a $Z'$ can also be used to extract \cite{ACLIII,leike}
its couplings
to quarks and leptons, yielding  information complementary to the
LHC. Such couplings can in turn allow one to determine the nature of the
underlying  extended gauge structure \cite{DLRSV,ACLIV}.

Since the photon couplings are only vector--like  and
the   $\ell$ couplings to $Z$ have the property
$\hat g_{L1}^\ell\simeq -\hat g_{R1}^\ell$  it turns out that
the  probes in the two--fermion
final state channels single out the $Z'$ leptonic couplings primarily in the
combinations $
\hat g_{L2}^\ell\pm\hat g_{R2}^\ell$.  To trace the
combinations of the normalized charges to which the  probes are sensitive, it
is
advantageous to choose either of the two  combinations to normalize
the charges.
 We choose the ${\hat g^\ell_{L2} - \hat g^\ell_{R2}}$ combination,
 which turns out to be a convenient choice for the typical models used in the
analysis. We then define  the following { four}
independent ``normalized''  charges:

\begin{equation}
P_V^\ell = \frac {\hat g^\ell_{L2} + \hat g^\ell_{R2}}
{\hat g^\ell_{L2} - \hat g^\ell_{R2}},
\ P_L^q = \frac {\hat g^q_{L2}}
{\hat g^\ell_{L2} - \hat g^\ell_{R2}},
\ P_R^{u,d} = \frac {\hat g^{u,d}_{R2}}
{\hat g^q_{L2}}.
\label{coup}
\end{equation}

\noindent
Their values  are given for the typical models in Table \ref{tnlc}.
In addition, the probes in the   two-fermion final state channels
 are sensitive to  the  following ratio of  an  overall gauge coupling strength
divided by the ``reduced'' $Z'$ propagator:
\begin{equation}
\epsilon_A= (\hat g_{L2}^\ell
- \hat g_{R2}^\ell)^2 \frac {g_2^2}{4\pi \alpha }
\frac {s}{M^2_{Z'} - s}.
\label{ep}
\end{equation}
 Here $\alpha$ is the fine structure constant.
Note again that the four normalized charges  (\ref{coup})
and $\epsilon_A$ (\ref{ep})
can be replaced with an equivalent set  by choosing
  ${\hat g^\ell_{L2} +\hat g^\ell_{R2}}$ to normalize the
couplings.

\begin{table}[hp]
\let\normalsize=\captsize
\caption{ Values of the couplings (\ref{coup}) and (\ref{ep}) for the
typical models
and  statistical error-bars as determined from probes
 at   the NLC ($\protect\sqrt s = 500$ GeV,
 ${\cal L}_{int}=20\ \hbox{fb}^{-1}$).  $M_{Z'} = 1$ TeV.
100\%\  heavy flavor tagging efficiency  and 100\%\ longitudinal polarization
of the
electron beam  is assumed for the first set of error bars, while the
error bars in   parentheses are for the probes without
polarization.}\label{tnlc}
\vskip.5pc
\footnotesize
\setlength{\tabcolsep}{5pt}
\renewcommand{\arraystretch}{1.3}
\begin{tabular}{|r|cccc|}
\hline
& $\chi$ & $\psi$ & $\eta$ & $LR$ \\ \hline
$P_V^\ell$ &$2.0\pm0.08\,(0.26)$&$0.0\pm0.04\,(1.5)$&
$-3.0\pm 0.5\,(1.1)$&$-0.15\pm 0.018\,(0.072)$\\
$P_L^q$&$-0.5\pm 0.04\,(0.10)$&$0.5\pm0.10\,
(0.2)$&$2.0\pm0.3\,(1.1)$&$-0.14\pm
 0.037\,(0.07)$\\
$P_R^u$&$-1.0\pm0.15\,(0.19)$&$-1.0\pm0.11\,
(1.2)$&$-1.0\pm0.15\,(0.24)$&$-6.0\pm1.4\,
(3.3)$\\
$P_R^d$&$3.0\pm0.24\,(0.51)$&$-1.0\pm0.21\,(2.8)$
&$0.5\pm0.09\,(0.48)$&$8.0\pm1.9\,(4.1)$\\
$\epsilon_A$&$0.071\pm0.005\,(0.018)$
&$0.121\pm0.017\,(0.02)$&
$0.012\pm0.003\,(0.009)$&$0.255\pm0.016\,(0.018)$\\\hline
\end{tabular}
\end{table}

One  should contrast the  above choice of the normalized couplings  with those
chosen for the LHC.
Recall that  couplings (\ref{tild}) probed by  the LHC, do not determine
 couplings (\ref{coup}) and (\ref{ep}) unambiguously.
In particular,  determination of $\gamma_L^\ell$, $
\tilde U$ and $\tilde D$ at the LHC
would yield  an eight-fold ambiguity   for the corresponding three
 couplings in (\ref{coup}) and (\ref{ep}).

The probes at the NLC  constitute
 the  cross sections  and corresponding asymmetries  in the
two-fermion  final state channels,  $e^+e^- \rightarrow f\bar f$.
 Due to the
interference of the $Z'$ propagator with the photon and the $Z$ propagators
such probes  are sensitive to the four normalized charges in  (\ref{coup}) as
well as  to  the parameter $\epsilon_A$ (\ref{ep}).
The tree-level expressions for such probes  can be
written explicitly in terms of
  seven generalized charges,  which are given in Ref. \cite{DLRSV} .

The analysis is based on the  following  probes:

\begin{equation}
\sigma ^{\ell},\   \ \ R^{had} = \frac {\sigma ^{had}}
{\sigma ^{\ell}},\ \ \ A_{FB}^{\ell}.
\end{equation}
In the case that  longitudinal polarization of the    electron
beam is available  there are additional  probes:
\begin{equation}
A_{LR}^{\ell, had},\ \ A_{LR,FB}^{\ell}\ .
\end{equation}
 Here $\sigma$, $A_{FB}$, $A_{LR}$ and $A_{LR,FB}$  refer to the
corresponding cross sections, forward-backward asymmetries, left-right
(polarization)
asymmetries and left--right--forward--backward asymmetries, respectively.
 The  superscripts $\ell$ and $had$
refer to all three leptonic channels (considering only
$s$-channel exchange for electrons) and  to all
hadronic final states, respectively.
The above  quantities help to distinguish among different
models \cite{DLRSV};  however,  they do not yield  information on all the
 $Z'$ couplings. In particular $\sigma^{\ell}$ and $A_{FB}^{\ell}$ probe
$\epsilon_A$  and the magnitude of $P_V^\ell$, but not  its
sign. On the other hand, $R^{had}$
 provides  additional information on one linear
combination of the normalized  quark couplings.
If polarization is available, $A_{LR}^{\ell}$ and
$A_{LR,FB}^{\ell}$  are excellent probes for $P_V^\ell$ (including its sign),
while
$A_{LR}^{had}$ yields  information on another linear combination of the quark
couplings.

LEP analyses \cite{LEP}\ show  that current $e^+e^-$ colliders allow for an
efficient  tagging of  charm ($c$) and bottom  ($b$) final states.
The  large momentum and the
nature of  the ($c,b$) lifetimes allow for flavor tagging by `flight'
identification.  At $LEP$ there
are three different methods for $b$ identification, based on lepton tagging,
event shape and lifetime tagging. On the other hand  $NLC$ detectors would be
(at least) as good as the corresponding $LEP$ ones. In addition, a larger
energy of jets at
the $NLC$ would imply a cleaner signature.   We therefore
assume  that at the $NLC$  an efficient tagging of the heavy flavors ($c,b,t$)
 would be available. This in turn provides an additional set of
observables:
\begin{equation}
R^{f}=\frac{\sigma^f}{\sigma^\ell}\ ,\ \ A^{f}_{FB} \ ; \ f=c,b,t\ ,
\label{rf}
\end{equation}
 and with polarization available:
\begin{equation}
A^{f}_{LR}\ ,\ \   A^{f}_{LR,FB}\ ;\ f=c,b,t\ ,
\label{af}
\end{equation}
 where the superscript refers to the corresponding
 heavy flavors.

\subsubsection{Determination of Gauge Couplings at the NLC}

We now study  how well one can determine the couplings defined in Section 5.2
at  the $NLC$.
The effects of a heavy $Z'$   far off-shell are
expected to be small and comparable to the electro-weak
radiative corrections \cite{DLRSV}\ . The latter ones are dominated by initial
state radiation, which can be greatly reduced by applying
a cut on the maximum photon energy to exclude $Z$
production. With such a  cut the tree-level expressions are
a reasonably good approximation to the different observables.
Since our present goal is to explore the sensitivity of the $Z'$ couplings, it
is sufficient to neglect the remaining radiative corrections. Of course, if a
new $Z'$ is actually discovered a realistic
fit should include full radiative corrections as well as
 experimental cuts and detector acceptances.

Only  statistical errors
for the observables  are included   and  error correlations for the input
parameters are neglected in the analysis.
In addition,  experimental cuts
and detector acceptances  are not included  either. The
results should thus  be interpreted as a limit on how
precisely  the couplings can be determined for each model for the given
c.m. energy and the integrated luminosity of the NLC.
 Realistic fits are expected to give larger uncertainties for the couplings.

\begin{figure}[htbp]
\let\normalsize=\captsize
\begin{center}
\centerline{\psfig{figure=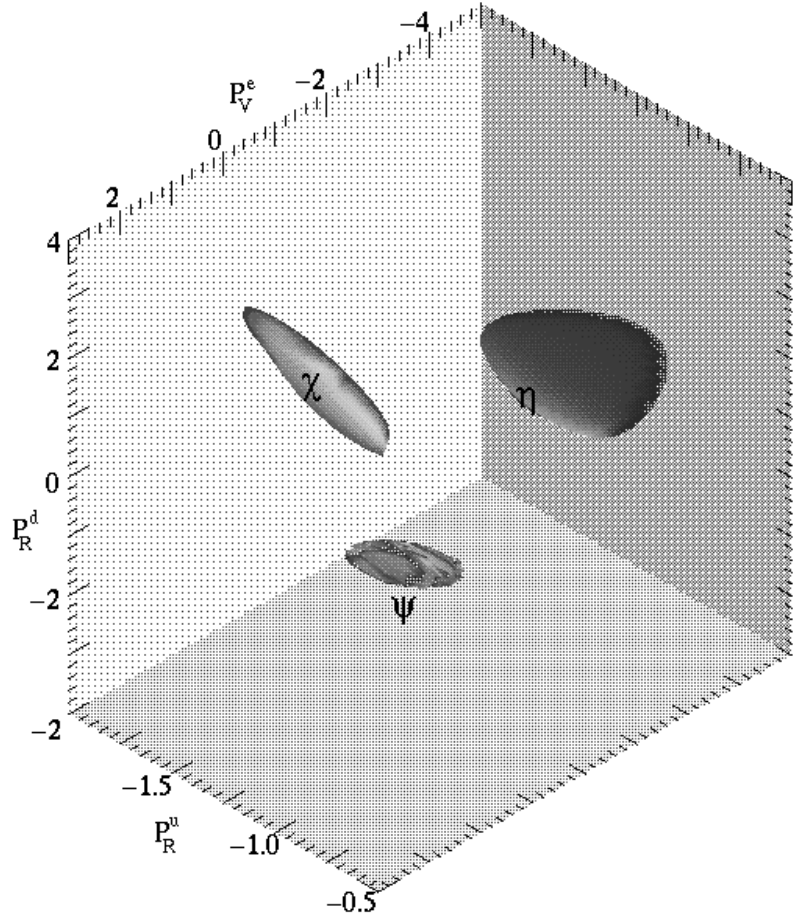,width=12.0cm}}
\begin{minipage}{12.0cm}
\caption[]{
90\%\  confidence level ($\Delta \chi ^2 = 6.3)$ regions for
the $\chi ,\psi $ and $\eta $ models  with $M_{Z'}=1$ TeV are plotted  on
$P_R^u$ versus $ P_R^d$ versus $P_V^\ell$ at the $NLC$.
}\label{cveticfig5}
\end{minipage}
\end{center}
\end{figure}

For $M_{Z'}=1$ TeV the couplings for the typical models  and the corresponding
statistical uncertainties are given in Table \ref{tnlc}.
In Fig.~\ref{cveticfig5},
90\%\ confidence level ($\Delta \chi^2=6.3$) regions  are given
in a three-dimensional plot of $P_R^u$ versus $P_R^d$ versus $P_V^\ell$  for
the $\chi$, $\psi$ and $\eta$ models (the $LR$ model is in a  different region
of  parameter space).
 100\%  efficiency for heavy flavor tagging   (probes) and  100\%
 longitudinal polarization  of the initial electron beam has been assumed.
 Relative error bars are about a factor of 2 larger than the corresponding
ones at the LHC.
The $Z'$  charges  can be determined to around $10-20\%$.

In the case of  smaller, say, 25\%,   heavy flavor tagging
efficiency   and  in the  case that
the electron beam polarization is  reduced to, {say},
50\%,  the determination of the couplings is poorer, though still useful
(see Ref.~\cite{ACLIII}.).

The diagnostic power of the $NLC$ for the $Z'$  couplings decreases
 drastically for
$M_{Z'}{\lower3pt\hbox{$\buildrel>\over\sim$}} 1$ TeV. {\it E.g.}, for
$M_{Z'}=2$  TeV, the uncertainties for the  couplings in the
typical models are 100\%, and thus a model-independent determination
of such couplings is difficult at the NLC.
For  $M_{Z'}\sim 2$  TeV, the uncertainties for the couplings in the
typical models  are too large to discriminate between models.

Given the $W'$ limits achievable at the upgraded Tevatron, it is
unlikely that $W'$ diagnostics will be possible at the 500~GeV NLC.
\clearpage

\subsection{Comparison of the Diagnostic Power of the NLC and the LHC}

The  couplings (\ref{tild})  that are  probed directly at the LHC
 are  not sensitive to the relative
signs of the $Z'$ charges. This in turn implies that
couplings (\ref{coup}), which are observed directly at the $NLC$,
 are probed with a few-fold ambiguity at the $LHC$.
In Table \ref{tcomp} we collect the errors
expected at the   $LHC$ for the three couplings  $P_V^{\ell}$, $P_R^u$ and
$P_R^d$.
 We again choose the typical models and $M_{Z'}=1$ TeV.
There is   an eight--fold ambiguity in determination of these couplings;
only the first value of $P_V^{\ell}$, $P_R^u$ and $P_R^d$
corresponds to the actual values of the typical models.
Note, however, that the  error-bars are
typically a factor of $\sim 2$  smaller than those at  the NLC (
compare  Tables  \ref{tnlc}  and \ref{tcomp}).

\begin{table}[htbp]
\begin{center}
\begin{minipage}{12.6cm}
\let\normalsize=\captsize
\caption{Values of three (out of four) couplings which  are
probed (indirectly) at the LHC.
The error-bars indicate how well these couplings
 can be measured at the LHC  ($\protect\sqrt s=14$ TeV and
${\cal L}_{int}=100$ fb $^{-1}$)  for the typical
models  with $M_{Z'}=1$ TeV. There is a two-fold ambiguity for
each of the couplings. Only the first number corresponds to the actual value of
the
coupling of the particular model.}\label{tcomp}
\vskip.5pc
\renewcommand{\arraystretch}{1.2}
\begin{tabular}{|c|cccc|}\hline
$ $ & $\chi$
& $\psi$ & $\eta$ &
$LR$ \\ \hline
$P_V^\ell$ & $ 2\pm 0.13 $ & $ 0\pm 0.03 $ &
$ -3\pm 0.15 $ & $ -0.148\pm 0.007 $ \\
$ $ & $ 0.5\pm 0.03 $ & $ \infty\pm \infty $ &
$ -0.333\pm 0.017 $ & $ -7\pm 0.36 $ \\
$P_R^u$ & $ \mp 1\pm 0.08 $ & $ \mp 1\pm 0.07 $ &
$ \mp 1\pm 0.04 $ & $ \mp 6.04\pm 0.27 $ \\
$P_R^d$ & $\pm 3\pm 0.09 $ & $ \mp 1\pm 0.11 $ &
$ \pm 0.5\pm 0.16 $ & $ \pm 8.04\pm 0.68 $ \\ \hline
\end{tabular}
\end{minipage}
\end{center}
\end{table}

In Fig. 6 we plot   90\%\ confidence level $(\Delta \chi ^2 = 6.3)$ regions for
the $\chi ,\psi $ and $\eta $ models  as
$P_R^u$ versus  $ P_R^d$  versus $P_V^\ell$ at the $LHC$.
While  the error-bars are small,  the figure displays a few-fold ambiguity for
 the value
of the couplings (\ref{coup}) (additional ambiguities are off the scale of the
plot).
 At the NLC the
error-bars are on the average
larger, but the   ambiguity in the value of the couplings is now removed.
Thus,  the LHC and the NLC are complementary and together
have the potential to uniquely determine
the couplings with small error-bars.

Finally,  we would also like to point out that the determination of
 $Z'$  couplings to   quarks and leptons  would in turn  allow
the determination of
the nature of the underlying  extended gauge structure \cite{ACLIV}.
As a prime example  one can choose the $E_6$ group. In this case
 two discrete constraints on  experimentally determined
  couplings have to be satisfied.  If so,
the couplings  would then uniquely determine
the two parameters, $\theta_{E_6}$ and $\kappa$,
which fully  specify the nature  of the $Z'$ within $E_6$.
If the $Z'$ is part of the $E_6$ gauge structure, then  for $M_{Z'}=1$ TeV
$\theta_{E_6}$ and $\kappa$
could be determined to around $10\%$  at the future colliders.  In
particular, one would be able to separate the  $LR$ symmetry breaking chains
from the  $SM\times U(1)'$ chains.
The NLC provides a unique determination of  the two
constraints  as well as of $\theta_{E_6}$ and $\kappa$,
though with
slightly  larger error bars than at the LHC. On the other hand,
since the LHC primarily determines  three out of four normalized couplings,
it  provides weaker constraints for the underlying gauge structure.  For more
details see Ref. \cite{ACLIV}.

\begin{figure}[htbp]
\let\normalsize=\captsize
\centering{
\centerline{\psfig{figure=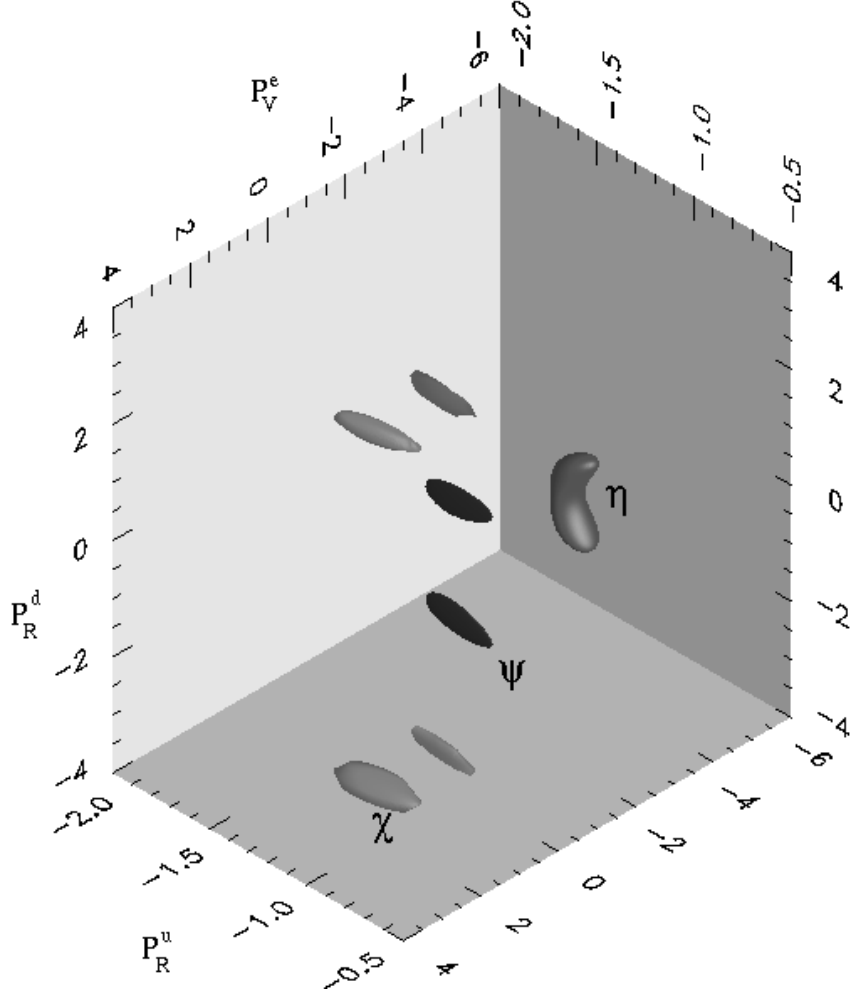,width=12.0cm}}
}
\caption[]{
90\%  confidence level ($\Delta \chi ^2 = 6.3$) regions for
the $\chi$, $\psi$, and $\eta$ models with $M_{Z'}=1$ TeV are plotted for
$P_R^u$ versus $P_R^d$ versus $P_V^\ell$ at the  $LHC$.
The figure reflects a few-fold  ambiguity in the
determination of these couplings at the $LHC$.}
\end{figure}


\section{Conclusions and Recommendations}

Among the facilities  operating in  the up-coming decade TEVATRON offers the
highest
discovery reach for new heavy gauge bosons  with masses up to 700--900
GeV range.

In the longer term,  future hadron colliders, {\it i.e.}, different
upgraded versions of the TEVATRON and the LHC, as well as a  high luminosity
$e^+e^-$  collider, {\it i.e.}, the NLC, would significantly improve limits
on the heavy gauge boson masses.
For the typical models such limits are in the 1--2 TeV region for the TEVATRON
upgrades, in the 4--5 TeV region for the LHC, and roughly
2-10$\times\sqrt{s}$ for the NLC with 50~fb$^{-1}$.

The LHC and  a high
luminosity 500 GeV $e^+e^-$ collider have discovery limits  for $Z'$ which
are in many cases  comparable.
Both  facilities have strengths and weakness.  The
limits obtained from the LHC are robust, in the sense that
they are obtained from a direct measurement with little background,
but  the sensitivity is dependent on the total width of the $Z'$, which
depends on
assumptions of the particle content of the model.
Limits obtained
for the NLC are indirect, based on statistical deviations from the Standard
Model, and are therefore more sensitive to  having the systematic
errors under control. However, they   do not depend on the unknown particle
content of the model.
With these caveats we would still consider,
for the purposes of discovering extra gauge bosons, the LHC
to be the machine of choice.

 We have also  explored  the diagnostic power of the LHC and  the
NLC  for a model independent determination of
$Z'$ couplings to quarks and leptons, once a $Z'$ is discovered.
In addition, the determination of such  $Z'$  couplings would in turn  allow
one to determine
the nature of the underlying extended gauge structure.
At the NLC,  efficient heavy flavor tagging
 and longitudinal polarization of the
electron beam provide probes in the two-fermion final state channels, which are
sensitive
 to  the  magnitude as well as the relative signs of  {\it all} the $Z'$
charges
to quarks  and leptons.  For $M_{Z'}
{\lower3pt\hbox{$\buildrel<\over\sim$}} $ 1 TeV, such
couplings would  be determined
to about 10--20 \%  for a class of typical models.
If polarization were not    available,
the determination of the  $Z'$  couplings  would  be  marginal, since the
error-bars
increase by a factor of 2--10.  Without heavy flavor tagging very
little  can be learned about the quark couplings.

The  LHC is complementary in nature; while
it primarily allows for the  determination  of  the magnitude
of only three out of four normalized couplings,
the corresponding errors  are  typically a factor of $\sim$2
  smaller than  those for the
NLC  for typical models with $M_{Z'}=1$ TeV.  In addition,
the LHC would measure $M_{Z'}$  directly and would
allow for a determination of an overall strength of the $Z'$ gauge coupling to
fermions. This is  in contrast to  the NLC which, for the fixed c.m.
energy, primarily determines only the ratio of an overall $Z'$ gauge coupling
strength and $M_{Z'}$.

The two machines possess complementary diagnostic power
for the model independent determination of the $Z'$ couplings to quarks and
leptons. In conjunction, they  allow for
determination of the $M_{Z'}$, an overall $Z'$ gauge coupling strength as well
as a  unique determination of   {\it all} the quark and lepton charges with
error bars in the 10--20\% range, provided $M_{Z'}\lsim 1$--2 TeV. In
addition,  at the LHC $W'$ diagnostics is  possible for $M_{W'}\le
1$--2 TeV.

A final observation is that if the DI-TEVATRON were running
prior to turning on of the LHC, the
non discovery of an extra gauge boson at the  DI-TEVATRON would
imply that the $Z'$ is too heavy to allow for diagnostic
study of its properties  at the LHC and/or the NLC.

\vskip 1.0cm

{\bf\noindent Acknowledgment}
\vskip 5mm
We  would like to thank  P. Langacker for  help with updating the results for
the global fits and other results as well as for the careful reading of the
manuscript. M.C. thanks
F. del Aguila for useful discussions and  S.G. thanks
T. Rizzo for helpful communications.

\end{document}